\documentclass[12pt]{article}
\pdfoutput=1
\usepackage{euscript,amsmath,amsthm,amsfonts,amssymb}
\usepackage{mathtools}
\usepackage{subfig}
\usepackage{float}
\usepackage[margin=1in]{geometry}
\usepackage{graphicx}
\usepackage{outlines}
\usepackage[dvipsnames]{xcolor}
\usepackage{jheppub}
\usepackage{cleveref}

\title{\boldmath The Crucial Calculation as a Motivating  Force  In Particle Physics}
\author{Howard J. Schnitzer}
\affiliation{Martin Fisher School of Physics, Brandeis University, Waltham, Massachusetts 02453, USA}
\preprint{BRX-TH-6668}
\emailAdd{schnitzr@brandeis.edu}
\abstract{Crucial experiments have a long history of contributions to progress in physics. Similarly, we claim that in the period roughly from 1955 to 1985 crucial calculations played a significant role in setting the agenda for  elementary particle physics. The highlights of the contributions of theoretical physics to the achievement of the standard model is emphasized.}

\begin{document}
\maketitle
\flushbottom

\section{ISSUES TO BE ADDRESSED}

\begin{enumerate}

\item  The development of elementary particle theory in the period from roughly 1955 to 1980 elucidates the manner in which theoretical physicists chose to follow a given line of research.

\item Elementary particle theory and quantum field theory in this period provides a good laboratory in which to ask the question since there were a great many directions investigated, although in some cases a direction may have only lasted a few years. Therefore it has the advantage of an overview of ``many generations" compressed into a short period of time [Fruit-fly analogy].

\item For the same reason, it may appear from the outside that particle physics was often a chaotic, ill-motivated discipline, subject to rapidly changing fashions. Again, from the outside it often seemed that some theorists behaved as ambulance chasers, and when arriving on the scene, a feeding frenzy occurred. The pack then seemed to move on quickly, leaving a few stragglers to pick over the remaining bones. In fact, it is my contention that the shifts in direction in particle theory were highly motivated, highly structured, and the result of the interplay of several complex factors.
        
\end{enumerate}
   
   \section{MOTIVATING FACTORS TO BE CONSIDERED}
   
   \begin{enumerate}

\item A crucial calculation may cause a large number of theorists to join a topic or leave a topic. There are examples of both factors.
\item  A crucial experiment can also be a strong motivating factor in change of direction.  Could it be that crucial calculations played a greater role than crucial experiments for the period in question?
\item The role of dominant figures is an extremely important, but not necessarily an overriding consideration. The most interesting situations occur when there exists more than one dominant figure at a given time, with divergent or conflicting views. Then how do members of the community choose their line of research? To ask this question necessarily places emphasis on the behavior of the highly competent, active theorists who are not the dominant figures, since given conflicting views, an active theorist must make a choice. The dominant personalities may be de-emphasized in making this choice.  If so, then how does a theorist choose a research direction?

Examples
   
     Crucial Calculations that motivated joining a direction

      \begin{enumerate}
\item Adler-Weisberger $\longrightarrow$ current algebra \cite{PhysRevLett.14.1051,PhysRevLett.14.1047}
\item Green –Schwartz anomaly calculation $\longrightarrow$ superstrings \cite{Green:1984sg}
\item Renormalizability of non-Abelian gauge theories  
$\longrightarrow$ QCD and the Standard Model  \cite{PhysRevD.5.823,THOOFT1972189}
\item Asymptotic freedom $\longrightarrow$ the Standard Model \cite{tHooft:1971qjg,PhysRevLett.30.1343,PhysRevLett.30.1346}
   \end{enumerate}

       A Crucial Calculation that motivated leaving a direction

             Coleman- Mandula $\longrightarrow$ leave  relativistic SU(6) \cite{PhysRev.159.1251}

       Crucial Experiments:

         That motivated a new direction
         
      \begin{enumerate}
\item  Discovery of Omega-minus $\longrightarrow$ SU(3) as a symmetry of strong interactions \cite{Samios:1980vh}
\item Weak neutral currents $\longrightarrow$ the standard model \cite{Erler_2013}
\item Discovery of J / $\psi$ $\longrightarrow$ acceptance of the quark model \cite{PhysRevLett.33.1404,PhysRevLett.33.1406}
   \end{enumerate}         
  
          That motivated one to leave a direction

                Deep inelastic scattering $\longrightarrow$ leave hadronic strings \cite{PhysRevLett.23.930}

         Baroque Explanations

                 Regge poles as fundamental to a description of strong interactions \cite{PhysRevLett.7.394}

         Strategic Retreat from strong interactions as fundamental, 
                 
                 Post 1960

\item  A Search for a Simple Explanation

                    When a particular line of research involves a formulation      
               which is ``too baroque", theorists will either abandon the line 
               completely, or seek to imbed it in a broader, simpler 
               description. Usually this does not involve a crucial calculation 
               (or experiment), but rather the desire of the community for a
               simple theory.

\item  Strategic Retreat

Sometimes a crucial calculation will close off avenues of research, without alternatives being available at the time.
The theory community must then reorganize its thinking along new lines. How is this done? Examples exist where this is a community effort, not that just of a major figure.

\item  Philosophical Underpinnings

            Some people do not leave a line of research
 
                  Foolhardy or courageous? This most frequently occurs with         
            “baroque“ theories. [S-matrix vs. string theories], or may           
            occur with simpler theories, but with little experimental support.
            It is very difficult to judge before and during a period of research.
            But things moved so rapidly in particle theory, some judgements
            could be made.   Philosophical underpinnings were not the issue
            for QM vs.Einstein, with QM accepted.  Chew in S-matrix theory  \cite{PhysRevLett.7.394},     
            had strong philosophical motivations, but the direction was 
            ultimately unprofitable. Green –Schwartz in string theory \cite{Green:1984sg} also                      
            has strong philosophical motivations, and that direction has   
            paid off in unexpected ways. Usually one must wait and see, 
            with possible surprises. In that context, string theory has   
            provided new insights into black hole issues.
          
\item  Theories with No Experimental Support   

     Why did theorists persist in pursuing theories with no visible (experimental) support, much to the puzzlement (or scorn) of their experimental colleagues?  They sought to embed baroque theories into simpler structures, and hoped to achieve clarification of first principles. [SUSY, SUGRA, superstrings, quantum gravity]. Hope springs eternal that a crucial calculation or experiment will establish the theory in some broad context (not necessarily definitive).

\item  Outcomes

                    Have these strategies employed by the particle physics been  
               successful   and economical of effort? I claim that there were 
              very few dead ends.   Essentially almost all of the mainstream                      
              ideas proposed have been incorporated or subsumed into later 
              theories.  However, the actual contribution of old ideas to
              newer theories may have sometimes taken unexpected form. Judgement of the efficiency of the effort is complex, but I
              believe that it has shown to be highly successful, well
              motivated, in a rapidly moving community.
        \end{enumerate}

          Specific Examples (more or less in chronological order, biased by
               personal experience)
               
               \begin{enumerate}

         \item Disaster [ A crucial calculation ends a subject]

               In September1955 I arrived at the University of Rochester as a
            1st year graduate student.  At that time, John Greene had just 
            defended his thesis attempting to explain the binding energy of 
            the deuteron as an n—p bound-state, bound by pion exchanges,
            using quantum field theory, as understood then \cite{jgreene}. Greene
            investigated pseudo-scalar and pseudo-vector pion-nucleon 
            couplings; pair suppression theories; intermediate coupling 
            theories, etc.; all to two loop order, i.e. up to 3 pion exchanges. 
            All these attempts to explain deuteron binding were failures. In
            retrospect he was studying the wrong problem, with the wrong
           degrees of freedom, and with the wrong methods. But that 
           would not become clear until roughly 10 years later.  At that time
           Rochester was a leading center of high energy physics, but there
           were parallels with Greene’s calculations found by others.
          To the best of my knowledge, this was the last field theoretic 
          calculation carried out in Rochester in strong interactions until
          many years later.
               It also marked the recognition that Nuclear Physics and Particle
          Physics were distinct disciplines.
           There were several possible issues raised by these failures. The 
         dominant possibilities considered were:
         \begin{enumerate}
\item The field theoretic description of the strong interactions was
         correct, but that perturbation theory was inadequate for the task.
\item Field theory, per se, might be incorrect at the short—
          distances probed by the nuclear forces; so that a fundamental
          reformulation of the theory was required.
                Nobody suggested that one was working with the wrong 
          degrees of freedom; as quarks and gluons appeared much later.
      \end{enumerate}
\item  Strategic Retreat and Reconstruction
   
        Two prevailing schools of thought; with two sets of dominant figures. These were:
           \begin{enumerate}
 \item  Dispersion relations (championed by Chew, and collaborators) \cite{smatrix} to organize, understand, and eventually predict the strong-interactions.
\item  Phenomenology of the weak interactions and the role of group theory (with a leading role played by Gell-Mann \cite{osti_4008239}, and later by Weinberg \cite{PhysRevD.8.605,PhysRevD.8.4482}, especially with current algebra).
       Both points of view involved treating physics at short-distances as a 
          black-box. Which was more fundamental, i.e. which one would 
          lead to a more fundamental reformulation? The betting in 1956 
          was that strong interactions were more fundamental. In any case, 
          theorists generally chose one or the other of the two approaches.
          To be more specific, let us look briefly at the choices faced by a 
          young theorist in choosing a direction for his research.

\item  Dispersion relations 
         The forward dispersion relations for pi-N scattering tested nothing
         but locality, microscopic causality, and unitarity (conservation of 
         probability). Causality required that field operators satisfy 
         space-like commutation relations. The forward pi-N scattering 
         amplitudes could rigorously be shown to satisfy a relation 
         analogous to the Kramers- Kronig relation of optics.  
         A test of the relation seemed to indicate failure of the relation.
                  A CRISIS
         The first research problem that I worked on resolved the question 
          in favor of the dispersion relation \cite{PhysRev.112.1802}. This was fundamental stuff, 
          right?  I cast my lot with the dispersion relation crowd. WRONG
         Dispersion relations did lead to S-matrix theory, Regge poles, but
         played a diminished role in the development of the standard 
         model.

 \item    Weak Interactions and Symmetries

           Using methods of group theory, one studied the weak 
        Interactions phenomenologically. For the most part, one used
        group theory to relate different (homologous) amplitudes,
        and searched for Lie groups to classify the known hadrons which
        came to populate an ever larger particle zoo.  At the outset, this
        seemed far removed from the 1st principles being probed by 
        dispersion theory.
                  Right? Wrong?
             This line of research gradually led to the V – A theory of the 
        weak interactions \cite{sudarshan1957proc,PhysRev.109.193}, SU(3) classification of the strong interactions \cite{osti_4008239}, 
        pions as Nambu-Goldstone bosons \cite{PhysRev.117.648,goldstone}, PCAC \cite{PhysRev.111.354}, soft-pions, current 
        algebra \cite{PhysRevD.8.605,PhysRevD.8.4482}, effective chiral Lagrangians, partons \cite{Feynman:1969wa,PhysRev.185.1975}, and the quark 
        model  \cite{osti_4008239}.  The effort to make the effective chiral Lagrangian  
        (which embodied all the information of current algebra) 
        compatible with the quark model \cite{GELLMANN1964214,zweigcern,PhysRevLett.13.598,PhysRev.139.B1006,PhysRevD.2.1285} led to the standard model
       of the strong interactions (quarks and gluons), to the
        (SU(2) x U(1)) electroweak theory \cite{PhysRevD.8.605,PhysRevD.8.4482,PhysRev.184.1625,Glashow:1961tr}, and finally the standard 
          model itself.
            There was no a priori way of anticipating the right direction in
        1956.  By the time current algebra had reached center stage in 
      the mid to late 1960’s Weinberg had become one of the dominant
      figures.
           At that time, there was little competition left from alternate
      points of view. The reconstruction had taken place, and there was
      a long period of verification of the standard model.
         \end{enumerate}
    \item   A Crucial Calculation vs. a Baroque Theory
  
          The crucial calculation which led most theorists to accept current
     algebra was the discovery and verification of the Adler-Weissberger
     relation.  The ingredients, reviewed above, were
     
          \begin{enumerate}
    
\item 	 the strong, electromagnetic, and weak currents were identical
(up to scale factors)
\item 	 the validity of equal-time current-algebra
\item 	 PCAC
\item 	the validity of the forward pi-N dispersion relations [ See above] 

      \end{enumerate}

(Note that similar dispersion relations played a parallel role in the development of other analogous sum rules testing current algebra, and later the parton model and quark model.)

Although Weinberg was a dominant figure in the development
       of current algebra, he did not lead theorists away from S-matrix 
      theory, Regge poles, hadron strings, etc. The success of current 
      algebra need not have led the abandonment of S-matrix theory,
      since there was no obvious conflict, at least at that time that 
      fashion shifted. What did?
      \begin{enumerate}
\item 	Challenged by the success of current algebra, Mandelstam \cite{PhysRev.184.1625}
       and others attempted to incorporate PCAC and current algebra
       into S-matrix theory, with no success. [A crucial calculation, as 
       a failure]. 
\item 	S-matrix phenomenology had become too baroque, too many Regge poles, and too many free parameters needed to fit data. How could this be a fundamental description? Apparently not.
\item 	Hadronic string theory had fundamental difficulties of principle; ghosts, tachyons, anomalies.
\item 	Hadronic string theory predicted that scattering cross-sections
would fall exponentially with momentum transfer, since the theory was “soft” at short distances. Deep inelastic scattering experiments showed that the fall-off was much more gradual (power-law) giving evidence of hard fundamental constituents at short-distances. This gave rise to the parton model and the quark-gluon model of the strong interactions \cite{salam}.
      \end{enumerate}
             
                 Thus, the shift in fashion from S-matrix theory to current 
                algebra and eventually to the standard model presented the
                 interplay of many complex issues for the individual 
                 theorist to consider:
  
        \begin{enumerate}    
\item 	a shift in dominant figures,
\item 	 a crucial calculation supporting current algebra (Adler-Weisberger),
\item 	a crucial calculation failure (Mandelstam and PCAC in S-matrix theory),
\item 	 a baroque phenomenology of the strong interactions vs. the desire for simplicity,
\item 	a crucial experiment deep inelastic scattering.

      \end{enumerate}
                    Later the definitive crucial experiment that convinced 
                 theorists that quarks were concrete degrees of freedom of the
                 underlying theory, and not mathematical bookkeeping devices
                 for the SU(3) classification of hadrons, was the discovery of 
                 the J/$\psi$ by the Ting and Richter groups \cite{PhysRevLett.33.1404,PhysRevLett.33.1406}.

\item  History Repeats?

       \begin{enumerate}
\item 	 Is the standard model the baroque theory of the present era? 
Too many free parameters. (masses, coupling constant, 
etc.).
\item 	 The effort to enlarge the theory so as to “simplify” the description of the standard model, in order to unify strong with electroweak and maybe gravity, has led to a wide variety of theories with no present experimental support; technicolor, supersymmetry, GUTS, supergravity, superstrings. However, although these theories are philosophically well motivated, which as we have remarked, is not a guarantee of success.
\item  Black hole unitarity is a fundamental issue.
\item  Cosmological constant a Pandora’s box?
\item  New theoretical attempt had roots in many lines of thought in particle theory. Many issues of principle need to be clarified, such as the compatibility of gravity with quantum theory. 
\item    Experimental support for “new” ideas?
       \end{enumerate}
 \end{enumerate}
 
      Summary
      
      \begin{enumerate}

\item 	 The post-war history of particle theory involved many complex and competing issues, leading to complicated decisions for a choice of research program by individual theorists.
\item 	 Directions changed rapidly as old lines of research became fully mature, or were closed off by crucial calculations or experiments. People generally did not stick with apparently unpromising directions.
\item 	 New directions were opened up by new crucial calculations, which were often not by the dominant figures.
\item 	 Dominant figures played an important, but not exclusive role in the particle physics community.
\item 	We were in an era where issues of principle and crucial calculations played a greater role than experimental information.  Indications are that this now may have shifted
back to a primary position for experiment. 
\item 	It is very difficult for the individual theorist to choose the most fruitful direction to pursue, based on available information at the time. Luck can be an important part of the choice.
\item 	 Outcomes of previous lines of thought are almost always to be found in present research areas.

\end{enumerate}

\section{CONTEMPORARY PARALLELS?}
Presently searches for “Beyond the Standard Model" is a theme which has not yet yielded concrete results. Clues may be coming from neutrino experiments, since neutrino masses point to “new physics". However, at the moment this search is dominated by experimental physics, not theory. LIGO, related gravitational wave searches, as well as significant results from astrophysics, have focused on the inclusion of gravity as necessary for a unified point of view. In this context, the interplay between black holes and unitarity is a prominent issue of theoretical physics, with a final unification not yet in sight. In my view, we are awaiting results from crucial experiments, with
crucial calculations playing a subsidiary role, in contrast to the “golden age" of elementary particle physics.

\acknowledgments
We are grateful to Isaac Cohen and Jonathan Harper for their aid in preparing the manuscript.

\vspace{\baselineskip}
\vspace{\baselineskip}
\noindent The choice of references is not intended to be comprehensive, but are chosen to be illustrative of the main issues of the text.

\bibliographystyle{JHEP}
\bibliography{Howard_Lecture}
\end{document}